\documentclass{jetpl}
\usepackage{epsfig,graphicx}
\twocolumn

\lat

\newcommand{\be}{\begin{equation}}
\newcommand{\ee}{\end{equation}}

\def\fun#1#2{\lower3.6pt\vbox{\baselineskip0pt\lineskip.9pt
\ialign{$\mathsurround=0pt#1\hfil ##\hfil$\crcr#2\crcr\sim\crcr}}}

\newcommand{{\SD}}{\rm SD}

\newcommand{\vex}{\mbox{\boldmath${\rm x}$}}
\newcommand{\vey}{\mbox{\boldmath${\rm y}$}}

\newcommand{\vesig}{\mbox{\boldmath${\rm \sigma}$}}

\newcommand{\vep}{\mbox{\boldmath${\rm p}$}}

\newcommand{\vez}{\mbox{\boldmath${\rm z}$}}

\newcommand{\veZ}{\mbox{\boldmath${\rm Z}$}}

\newcommand{\veH}{\mbox{\boldmath${\rm H}$}}
\newcommand{\veE}{\mbox{\boldmath${\rm E}$}}

\newcommand{{\Mc}}{\mathcal{M}}
\newcommand{\llan}{\langle\langle}
\newcommand{\rran}{\rangle\rangle}
\newcommand{\lan}{\langle}
\newcommand{\ran}{\rangle}

\title{Electric charge in the  stochastic  electric field  }
\rtitle{Electric charge in the  stochastic  electric field }
\sodtitle{Electric charge in the  stochastic  electric field }

\author{  Yu.\,A.\,Simonov$^+$ \thanks{simonov@itep.ru}}
\rauthor{Yu.\,A.\,Simonov}
\sodauthor{Simonov}

\address{$^+$ Institute of Theoretical and Experimental
Physics\\ 117218, Moscow, B.Cheremushkinskaya 25, Russia}

\dates{04 May 2016}{*}

\abstract{The influence of electric    stochastic fields on the relativistic charged particles  is  investigated in the  gauge invariant path integral formalism. Using the  cumulant expansion one finds the exponential relaxation of the charge Green's function both   for spinless and Dirac charges.}

\PACS{11.15.Tk, 11.15.-q, 12.20.-m}

\begin{document}

\maketitle

\section{ }
Stochastic electromagnetic fields and specifically stochastic electric fields
play important role in many areas of physics and technology,  from
extragalactic fields and  cosmology \cite{1,2}, to  molecular physics \cite{3}
medicine (e.g. tomography),  accelerator physics \cite{4}; for a general theory
 see \cite{5}.
 Very intensive electric and magnetic fields occur in the heavy ion collisions
 \cite{5'} with possible stochastic component.

The treatment of the dynamics of the system in a stochastic field can be done
on the general grounds in the   formalism of the (world-line) path integrals
\cite{10,11}, which is applicable to any system and in principle, to  any
field, see \cite{6} for a recent development. In particular it was shown  in
\cite{10,7,8}, that stochastic colorelectric  fields give  rise to the basic
phenomenon of confinement. In this derivation it was essential, that the
correlation length $\lambda$ of  stochasticity is much smaller, than other
ranges in the theory, e.g. the radius of orbital motion etc.  Another important
feature was the Euclidean character of colorelectric fields, which ensures the
real confining potential.

In this paper we apply the path integral approach to the relativistic field
theory in the external  real  electric fields, and discuss the  simplest
situation, when the correlation length $\lambda$ (correlation time $\tau$) is
much smaller than other lengths (periods) in the problem, denoted by $\Lambda$,
i.e. the free path (time interval) of the charge between collisions, the  range
of constant potential  (period  of orbital motion) on the atomic or molecular
level.

In this case the effect of stochasticity enters as an additional term in the
total Hamiltonian; it is imaginary for stochastic electric fields. In the
opposite case, $\lambda >\Lambda$, the stochastic fields are acting on the
defined motion of the particle, and the  formalism becomes more complicated.

\section{ }
 We start with the Green's function of one spinless charged relativistic
 particle in the  external electromagnetic  field, which   will be treated as
 stochastic in the cumulant formalism \cite{9}.  To this end we write down the
 path integral form  for the Green's function, given by the
 Fock--Feynman--Schwinger (FFS) path integral \cite{10,11}

\be G(x,y) =(m^2 -D^2_\mu)^{-1}_{xy} = \int^\infty_0 ds D^4 z e^{-K} \Phi
(x,y),\label{1}\ee
where
 \be \Phi(x,y) =\exp (ie\int^x_y A_\mu(z)
dz_\mu),\label{2}\ee
 \be K=\int^s_0 [ m^2 +\frac14 \left(
\frac{dz_\mu}{d\tau}\right)^2] d\tau.\label{3}\ee

$ A_\mu(z)$ can include both stochastic and nonstochastic (e.g. Coulombic)
components, but we shall start for simplicity with the stochastic part only. It
is convenient to work out the final expressions in the Euclidean space-time,
and then go over to the Minkovskian space-time, however  finally,  all time
fluctuations and the fields  should considered in the real (Minkovskian) time.

Choosing the virtual energy $\omega$ instead of the proper time $s$ as a
variable, $ s=\frac{T}{2\omega}, T=x_4-y_4$, one has \be \int^\infty_0 ds
(D^4z) \Phi(x,y) e^{-K} = T \int^\infty_0 \frac{d\omega}{2\omega^2} (D^3
z)_{_{\vex\vey}} e^{-K(\omega)} \Phi (x,y)\label{4}\ee where \be K(\omega) =
\int^T_0 d t_E \left( \frac{\omega}{2} + \frac{m^2}{2\omega} +
\frac{\omega}{2}\left( \frac{d\vez}{dt_E}\right)^2\right)\label{5}\ee

To define stochastic fields in a gauge-invariant way  as averages of field
strengths one considers any trajectory among the  integral paths and combine it
with a fixed shadow trajectory, which is needed to complete the closed contour
$C$. This helps to express the  vector potential $A_\mu(z)$ in terms of the
field strength, e.g. via the representation  \cite{13} \be A_\mu (z) = \int^z_0
F_{\nu\mu} (u) \alpha(u) du_\nu,\label{6}\ee and $\Phi(x,y)$, which is  now a
part of the Wilson loop $C(x,y)$,including the charge trajectory in $\Phi
(x,y)$  and the shadow trajectory,  can be rewritten in the form \
\begin{multline}
W(x,y) = \exp\left(\int_{C(x,y)} ie A_\mu (z) dz_\mu\right) =\\
 \exp (ie \int ds_{\mu\nu}(u) F_{\mu\nu} (u)),
\label{7}
\end{multline}
where $ds_{\mu\nu}$ is the surface element in
the area inside $C(x,y)$, $ ds_{\mu4} = du_\mu dz_4, $  and  $ dz_4$ is the
Euclidean time element $dz_4=idz_0$ ,

At this point it is important to define the stochastic process \cite{5,9},
namely, to take into account, that $\Phi(x,y) \to W(x,y)$ enters as an average
value in the averaging procedure, which is characterized by the  gauge
invariant correlators of the field strengths, e.g.
\begin{multline}
e^2 \lan F_{\mu\nu} (x)F_{\lambda\sigma} (y)\ran =\\
\frac12 \left( \frac{\partial}{\partial w_\mu}(w_\lambda D^{(2)} (w)) \delta_{\nu\sigma} + \frac{\partial}{\partial w_\nu} (w_\sigma D^{(2)} (w) \delta_{\mu\lambda})\right) 
\label{8}  
\end{multline}
 and similarly for the third and higher powers of $F_{\mu\nu}$. Here $w_\lambda = x_\lambda -
y_\lambda$, and  $D^{(2)}$ is a scalar function of  its arguments, $D^{(2)} (w)
\to D^{(2)} (w_i^2+ w^2_4)= D^{(2)} (w^2_i-w^2_0).$

One can now express the average value $\lan W \ran$ in terms of  cumulants of
field correlators \cite{6,9}, assuming that $\lan F_{\mu\nu}\ran \equiv 0$,
\begin{multline}
\lan W(x,y) \ran =\\
\exp \sum \frac{(ie)^n}{n} \int ds_{\mu_1\nu_1} (1) ...ds_{\mu_n \nu_n} (n) \llan F_{\mu_1\nu_1} (1) ... F_{\mu_n\nu_n} (n) \rran
\label{9}  
\end{multline}
 where $\llan F...F\rran$ are cumulants, defined as follows
\be \llan F_{\mu\nu_1} (1) F_{\mu_2\nu_2} (2) \rran  \equiv \lan F_{\mu_1\nu_1}
(1) F_{\mu_2\nu_2} (2)\ran\label{10a}\ee
\begin{multline}
\llan F_{\mu_1\nu_1} (1) F_{\mu_2\nu_2} (2)  F_{\mu_3\nu_3} (3) F_{\mu_4\nu_4}(4) \rran  =\\
 \lan F  (1) F (2) F  (3) F  (4) \ran - \lan F  (1) F (2)\ran  \lan F  (3) F (4) \ran -\\
 \lan F(1) F(3) \ran \lan F(2) F(4) \ran - \lan F(1) F(4) \ran \lan F(2) F(3) \ran .
\label{10}  
\end{multline}
 see \cite{9} for more
details.

 In principle, as  it  was shown in \cite{6}, one  can express   QED or QCD
 dynamics in terms of correlators , e.g. for the Coulomb interaction (the field
 correlator of the  photon exchange) the corresponding correlator is
 $D^{(Coul)}(w)$,
 \be D^{(2)(Coul)} (w) =\frac{4\alpha}{\pi(w^2_i+w^2_4)^2},\label{12}\ee
 and its contribution to $\lan W\ran$ is  \be \lan W(x,y) \ran =
 \exp \left( i \int \frac{e^2d z_0}{r(z_0)}\right)\label{13}\ee
 where $r(z_0)$ is the  distance between the point  $ \vez(z_0)$ on the
 trajectory of the particle, and  the point $\veZ (z_0)$ on the shadow
 trajectory,  which can refer to the opposite charge companion of our particle.
 In the case when one can neglect the Coulomb interaction of the particle with
 all neighbors, the correlator (\ref{12}) is absent, and  we assume, that all
 stochastic correlators are fast decreasing,  both in time and in space variables,
 so that the average $\lan W \ran$
 factorizes into a product of averages separately for a given charge trajectory
 and a shadow trajectory, which will  not be of interest to us.

 As a result the average $\lan W \ran$ in (\ref{9}) contains integrations
 $ds_{\mu_i\nu_i}$ in the vicinity of the trajectory $z_\mu(\tau) \to z_\mu(t)$.

 Keeping only the term $n=2$ in (\ref{9}), one obtains
  in the  exponent of (\ref{9})

\begin{multline}
\Gamma_E T_4= \frac{e^2}{2} \int d s_{\mu_1\nu_1} (u) ds_{\mu_2\nu_2} (v) \lan F_{\mu_1\nu_1} (u) F_{\mu_2\nu_2} (v)\ran =\\
 \frac{1}{4} \int   ds_{\mu_1\nu_1} (u) ds_{\mu_2\nu_2} (v)\left( \frac{\partial}{\partial w_{\mu_1}} (w_{\mu_2} D^{(2)} (w)) \delta_{\nu_1\nu_2} + \right.\\
\left. \frac{\partial}{\partial w_{\nu_1}} (w_{\nu_2} D^{(2)} (w))
\delta_{\mu_1\mu_2}\right).
\label{16}   
\end{multline}

For the stochastic electric fields  $F_{\mu_1 4} , F_{\mu_2 4} = i E_{\mu_1},
iE_{\mu_2}$  one obtains, assuming that the surface in the loop $C(x,y)$ lies
in the $x_1, x_4$ plane,  $ 0\leq u_1, v_1 \leq R; ~0\leq u_4, v_4 \leq T_4$

\begin{multline}
\Gamma_E =\frac{1}{2} \int^R_0 du_1 \int^R_0 dv_1 \int^{T_4}_{-T_4} dw_0 \left( \frac{\partial}{\partial w_{ 1}} (w_{1} D^{(2)} ) + \right.\\
\left. \frac{\partial}{\partial w_{0}} (w_{0} D^{(2)})\right).\label{17}  
\end{multline}
We assume, that $D^{(2)}$ falls off fast for large $|w_4|=|w_0|$, so that the last factor
in (\ref{17}) vanishes, while the first term yields for the charge trajectory
(a similar answer results for the background one) \be \Gamma_E =
\int^{T_0}_{-T_0}  {dw_0} \int^R_0 dx \cdot x D^{(2)} (x, w_0).\label{18}\ee

Note, that $D^{(2)} (0,0)\geq 0$ for real electric fields.  As a result the
leading quadratic correlator leads to the following result for $\lan W\ran$\be
\lan W (x,y) \ran = \exp (- \Gamma_E T_0), ~T_0 =x_0-y_0>0.\label{19}\ee Note,
that we have used in (\ref{17}) both temporal and spacial stochastic
correlations, present in the correlator $D^{(2)}$ in (\ref{18}).

Inserting (\ref{19}) into (\ref{4}) in the place of $\Phi(x,y), \lan \Phi(x,y)
\ran \to \lan W (x,y) \ran$, one obtains $ (T_4 \equiv  T = i T_0)$ \be \lan
G(x,y)\ran_A = \sqrt{\frac{T_4}{8\pi}} \int^\infty_0 \frac{d\omega}{
\omega^{3/2}} (D^3 z)_{\vex\vey} e^{-K(\omega)+ i  \Gamma_E T_4}.\label{19a}\ee

 Doing  the $ (D^3 z)$ integration as in \cite{6}, one introduces the Hamiltonian $H(\omega)$
   \be \lan G(x,y)\ran_A =
\sqrt{\frac{T_4}{8\pi}} \int  \frac{d\omega}{ \omega^{3/2}} \lan \vex |
e^{-(H(\omega)-i   \Gamma_E) T_4}|\vey\ran,\label{20}\ee where $H(\omega)$ is,
\be H(\omega) = \frac{\vep^2+m^2}{2\omega} + \frac{\omega}{2} +
V_Z(r)\label{21}\ee and we have included the potential $V_Z(r)$ in the case,
when our particle is inside an atom or a hadron.

 Integrating over $d\omega$ by the  stationary point method as in \cite{6}, one arrives
at the final form,
 \be\lan  G_Z (x,y) \ran  = \frac{1}{2m} \sum_n \varphi_n^* (\vex)
 \varepsilon^{-i \varepsilon_n T_0 -  \Gamma_E T_0} \varphi_n (\vey),\label{23}\ee
 where $\varphi_n(x)$ is atomic bound wave function, while \be \varepsilon_n
 =M_Z + m \sqrt{1-\frac{(Z\alpha)^2}{n^2}}.\label{24}\ee

 So far the spinless particles  were considered. Now, for particles with the
 spin 1/2 one should add in  the exponent of (\ref{7}) the term (see
 \cite{10,11}),
 \be i e \int \sigma_{\mu\nu} F_{\mu\nu} \frac{dt_4}{2\omega}; ~~
 \sigma_{\mu\nu } F_{\mu\nu} =\left( \begin{array}{ ll} \vesig \veH&  i \vesig
 \veE\\ i \vesig
 \veE &\vesig \veH\end{array}\right).\label{23a}\ee

 Keeping only Gaussian (quadratic) correlator of $\veE (x,t)$, and going to the
   Minkovskian time, one obtains in the exponent of (\ref{16}) for the
 spinor case
 \be \Gamma_E \to  \Gamma_E  + \Gamma_E^\sigma \hat 1, \Gamma_E^\sigma =
 \frac{1}{8m^2} \int \lan E_i (z_0) E_i (z'_0)\ran d (z_0 - z'_0).\label{24}\ee
Here $\hat 1$ is the  unit $4\times 4$ Dirac matrix. Thus one can see, that
spin-dependent  forces, bring about additional relaxation of the Green's
function.

\section{} We have shown above,  that high-frequency stochastic electric fields
with the correlation length (time) smaller than  other periods of time in the
particle motion, create the relaxation of the particle signal. The  relaxation
widths $\Gamma_E, \Gamma_E^\sigma$ are proportional to the  quadratic cumulant
of the field strength. One can check, that higher correlators do not change
qualitatively the situation. It is clear, that the weakening of the signal of
the given total energy (momentum) is caused by the   redestribution of energies
(momenta) in the  combined collective signal as  can be  seen by expanding
fluctuating field $E_i(x,t)$ in the Fourier series, and this should be
described in the framework of the general stochstic approach \cite{2,5}. One
example where these processes may be important, is the behavior of the
stochastic quark-gluon ensemble in the process of heavy ion collisions, where
occur electric and magnetic fields of high intensity \cite{5'}.

Consider now any macroscopic current $J(x,t)$, which is a statistical or coherent sum of elementary currents
with the Green's functions given by Eq.(1).It is clear,that the correlator  of the currents $\langle J(x,t)J(x',t') \rangle$
will be proportional to  $exp(-\Gamma_E |t -t'|)$, implying that the high-frequency stochastic background creates the universal relaxation of the signal,since $\Gamma_E$ in (16) depends only on the stochastic source characteristics,namely on its high-frequency tail.
 It is remarcable from the general physical point that the same stochasic process with Euclidean colorelectric
fileds in QCD creates confinement and hence all our world,while in QED an analogous stochastic process
simply yields a universal damping of any signal.

The author is grateful to   M.A.Andreichikov and   B.O.Kerbikov  for
discussions. The  financial support of the Russian  Science Foundation via the
grant  16-12-10414 is gratefully  acknowledged.


\begin{thebibliography}{99}
%
\bibitem{1} E.Calzetta, A.Kandus, Phys. Rev. {\bf D 89}, 083012 (2014).

\bibitem{2} E.Calzetta and B.L.Hu, Nonequilibriun Quantum Field Theory,
Cambridge university Press, Cambridge (2008).

 \bibitem{3}Y.B.Band and Y.Ben-Shimol, Phys. Rev. {\bf E 88}, 042149 (2013).

\bibitem{4} R.Martorelli, G.Montani and N.Carlevaro, Mod. Phys. Lett. {\bf
A 31(2)} , 1650005 (2016).



\bibitem{5} N.G.van Kampen, Stochastic processes in Physics and Chemistry,
Elsevier, 1997.

%
\bibitem{5'}D.E.Kharzeev, L.D.McLerran and H.J.Warringa, Nucl. Phys.A {\bf 803}, 227
 (2008); V.Skokov, A.Illarionov and V.Toneev, Int. J. Mod. Phys.A {\bf 24},
 5925 (2009).

 \bibitem{10} Yu.A.Simonov, Nucl. Phys. {\bf B 307}, 512 (1988).

\bibitem{11} Yu.A.Simonov and J.A.Tjon, Ann. Phys.  {\bf 228}, 1 (1993), ibid {\bf 300}, 54 (2002).

\bibitem{6}Yu.A.Simonov, Phys. Rev. {\bf D 88}, 025028 (2013), ibid. {\bf D 90}, 013013 (2014).



\bibitem{7}

  H.G.Dosch, Phys. Lett. {\bf B  190}, 177 (1987);\\
  H.G.Dosch and Yu.A.Simonov, Phys. Lett.{\bf  B 205}, 339 (1988).
\bibitem{8}Yu.A.Simonov, Phys. Uspekhi, {\bf 39}, 313 (1996).

\bibitem{9}
N.G.van Kampen, Phys. Rep. {\bf C 24}, 171 (1976); Physica {\bf 74}, 215
(1974).








\bibitem{13} I.I.Balitsky, Nucl. Phys. {\bf B 254}, 166 (1985).

\end{thebibliography}
\end{document}